\documentclass[letterpaper,12pt]{article}
\pdfoutput=1
\usepackage[letterpaper,margin=1in]{geometry}
\usepackage{graphicx}
\usepackage{amssymb}
\usepackage{amsmath}
\usepackage{siunitx}
\usepackage{xcolor}
\usepackage[colorlinks=true,
            urlcolor=blue,
            linkcolor=blue,
            citecolor=blue]{hyperref}
\usepackage{lineno}
\usepackage[T1]{fontenc}
\begin{document} 

\begin{center}
\Large\textbf{%
Canonical form of a deformed Poisson bracket spacetime
}
\end{center}

\centerline{Douglas M. Gingrich}

\begin{center}
\textit{%
Department of Physics, University of Alberta, Edmonton, AB T6G 2E1 Canada\\
\smallskip
TRIUMF, Vancouver, BC V6T 2A3 Canada
}
\end{center}

\begin{center}
e-mail:
\href{mailto:gingrich@ualberta.ca}{gingrich@ualberta.ca}
\end{center}

\centerline{\today}

\begin{abstract}
\noindent
The general uncertainty principle applied to gravity can be implemented as a 
set of modified Poisson brackets in the canonical formalism.
As such, the theory is not canonical and the resulting
equations of motion do not lead to a covariant metric.
We construct a Hamiltonian that when applying the usual canonical formalism
gives a closed algebra and equations of motion that result in the original
metric obtained by using distorted Poisson brackets. 
The resulting theory is thus rendered canonical and covariant.
We then covariantly couple scalar matter and dust to the modified gravity to
allow the study of dynamics.
\end{abstract}
\section{Introduction}

One of the grand challenges in physics is to reconcile quantum mechanics and
gravity.
A theory of quantum gravity is thought necessary to understand extreme
conditions like those inside black holes and at the Big Bang. 
The most promising, or at least most popular, approaches to date have been
string theory and loop quantum gravity, both of which are not entirely
satisfactory. 

To get a hint of which direction to go, it becomes necessary to explore ideas
at a more phenomenological level that modify general relativity to achieve the
desired goals that we believe a full quantum theory will solve.
One necessity of these models, as a bare minimum, is the resolution of
singularities that plague general relativity.
The models often also incorporate plausible feature we believe a full quantum
theory of gravity could possess.

There are a class of long-standing modified gravity theories which take the
approach of modify the Einstein-Hilbert action of general relativity.
Theories such as $f(R)$ gravity~\cite{10.1093/mnras/150.1.1},
scalar-tensor gravity~\cite{PhysRev.124.925}, and higher-order curvature
invariants~\cite{10.1063/1.1665613}, to name just three, have had varying
success and are well established fields of study.  
As they start from the Lagrangian formalism they are necessarily covariant.

While these studies are of fundamental importance, one should also consider
other types of modified gravity beyond simply adding terms to the action.
These ideas, often being generated heuristically, are not necessarily
covariant~\cite{Bojowald:2020unm} and can lack a connection to underlying
theories. 
But at the same time can be highly effective at singularity resolution and
incorporating other desirable features we believe a quantum theory of gravity 
could allow.

Since the uncertainty principle is foundational to quantum mechanics, one might
posit incorporating a modified version of the uncertainty
principle~\cite{Maggiore:1993rv,Kempf:1994su,Scardigli:1999jh,
Bosso:2019ljf,Bosso:2023aht,Melchor:2023rqd,YupanquiCarpio:2024asq}
into gravity as a method for capturing some quantum aspects of the theory.  
One approach is to modify the Poisson bracket in the canonical formalism of
gravity to incorporate some general uncertainty principle (GUP) considered as
an extension of quantum
mechanics~\cite{Blanchette:2021vid,Bosso:2020ztk,Rastgoo:2022mks,Bosso:2023fnb}.  

We consider the approach taken in~\cite{Fragomeno:2024tlh,Gingrich:2024mgk}.
After modifying the Poisson brackets between conjugate variables to represent a
GUP in the black hole interior, solutions to the equations of motion of the
triad are found. 
From these solutions, the interior metric is constructed.
The interior metric is then analytically extended to the full spacetime by
switching the timelike and radial spacelike coordinates.
The singularity is resolved, and the correct classical and asymptotic limits
are obtained. 
Since the Poisson brackets have been distorted, the theory
is not canonical and is not covariant (see Appendix~\ref{sec:appA}).
Here we remedy this situation.

To connect the GUP spacetime to the Hamiltonian canonical formalism of
gravity, we consider the most general Hamiltonian constraint up to second order
in the derivatives and quadratic in the first-order derivatives of the
phase-space variables.
It must form a closed hypersurface deformation algebra with the diffeomorphism
constraint, and thus lead to a dynamical flow covariantly defining a spherical 
symmetric four-dimensional geometry and hence metric on the
manifold~\cite{Alonso-Bardaji:2023vtl,Bojowald:2023xat}.    

Starting from the GUP inspired spacetime, we apply a method that constructs
back to the Hamiltonian, such that its dynamical flow covariantly reproduces
the given GUP metric.   
The Hamiltonian is defined as a linear combination of the standard radial
diffeomorphism constraint and a deformed Hamiltonian constraint with respect to
general relativity which incorporates the GUP corrections.
The formalism is covariant.
The solution to the constraint equations in different gauges will give the
same geometry, though in different coordinate charts.

The Hamiltonian construction provides a dynamical theory.
Thus allowing a consistent and covariant coupling to matter fields.
The coupling to matter allows the calculation of its evolution using test
fields on the spacetime, as well as the backreaction of the matter fields on
the geometry. 
It also allows a determination of whether this particular black hole model may
emerge as the end result of dynamical collapse.

An outline of this paper is as follows.
In Sec.~\ref{sec:phasespace}, we summarize the Hamiltonian formalism of
classical gravity thus defining the terms and notation that will subsequently be
used. 
We revisit the Poisson brackets leading to the GUP inspired spacetime in
Sec.~\ref{sec:gup}.
The metric is transformed to correspond to a form that can be converted to the 
Hamiltonian constraint.
In Sec.~\ref{sec:hamiltonian}, we write down the Hamiltonian constraint for the
GUP inspired spacetime and obtain the equations of motion of the emergent
gravity phase-space variables.
In Sec.~\ref{sec:gauge}, we consider the static and homogeneous gauges and
obtain the same line element that resulted from a distorted Poisson algebra.
By construction the Hamiltonian leads to a covariant theory.
We couple matter to the metric in Sec.~\ref{sec:matter}.
The deformed Poisson bracket in the static gauge is considered in
Appendix~\ref{sec:appA}.
Appendix~\ref{sec:appB} gives more details about the calculation of the
hypersurface deformed algebra.
Appendix~\ref{sec:appC} derives the scalar equation of motion in a
spherically symmetric spacetime using the Hamiltonian formalism and shows it to
be identical to that derived in the same spacetime using the action
(Appendix~\ref{sec:appD}). 
Dust is also coupled to the modified gravity.
The paper finishes with a summary and comments on how the results can, and
will, be used in future work to study gravitational perturbations and collapse. 

\section{Phase-space description\label{sec:phasespace}}

In this section, we review the canonical gravity formalism and set the
notation which we will use.
We follow~\cite{Alonso-Bardaji:2023vtl} and in particular the notation
of~\cite{Belfaqih:2024vfk}. 
The phase-space coordinates consist of four fields $K_x, E^x, K_\varphi$, and
$E^\varphi$, where each is a function of time $t$ and a radial coordinate $x$
in the spherically symmetric manifold.
In the classical theory, the momenta $E^x$ and $E^\varphi$ are components of
the densitized triad\footnote{A triad is the space part of a tetrad.
They are orthogonal vector fields that can be used to construct the spatial
part of the matrix. Triads are densitize by multiplying the triad by the
square root of the determinant of the spatial metric.},
while the configuration variables $K_x$ and $K_\varphi$ are directly related to
the extrinsic curvature components $\mathcal{K}_\varphi K_\varphi$ and
$\mathcal{K}_x = 2K_x$. 

The symplectic structure\footnote{A canonical symplectric structure provides a
geometrical form for Hamiltonian mechanics and phase space.
The symplectic structure and Poisson brackets are canonically connected, thus
linking the geometrical structure to the algebraic structure.}
is chosen to be canonical, such that for a give $t$ the only nonvanishing
Poisson brackets are 

\begin{equation}
\{ K_x(t,x_1), E^x(t,x_2) \} = \delta(x_1-x_2)
\quad\text{and}\quad
\{ K_\varphi(t,x_1), E^\varphi(t,x_2) \} = \delta(x_1-x_2).
\end{equation}

The diffeomorphism constraint arises from the requirement that
the theory is invariant under spatial coordinate transformations.
It ensures that the physical states of the theory are independent of the choice
of spatial coordinate on the three-dimensional hypersurface.
In the ADM formulation of general relativity, the diffeomorphism constraint is
obtained by varying the action with respect to the shift vector
(see~\cite{Bojowald_2010}). 
The resulting diffeomorphism constraint is

\begin{equation}
\mathcal{H}_x = - K_x (E^x)^\prime + E^\varphi K_\varphi^\prime,
\end{equation}
where primes represent derivatives with respect to $x$.
The Hamiltonian constraint $\mathcal{H}$ is given
in~\cite{Alonso-Bardaji:2023vtl} and an equivalent version
in~\cite{Bojowald:2023xat}. 
Since we do not directly use either of these Hamiltonians, we do not
reproduce them here.

Since the phase-space variables are fields, the Poisson
brackets are not strictly an algebra but a distribution.
To avoid using distributions, one usually smears the fields or functionals of
them.
Since the Hamiltonian constrains are a function of the fields, we can integrate
over them times an arbitrary function, including the spatial determinant
since we are on a curved manifold.
One thus sees the utility of densitizing the triad.
We define the usual smeared forms for the Hamiltonian for arbitrary
function $s$:

\begin{equation}
H_x[s] = \int s \mathcal{H}_x dx
\quad\text{and}\quad
H[s] = \int s \mathcal{H} dx.
\end{equation}
The hypersurface deformation brackets for spherical symmetry are

\begin{align}
\{ H_x[s_1], H_x[s_2]\} &= H_x[s_1 s_2^\prime - s_1^\prime s_2],\label{eq:b1}\\
\{ H_x[s_1], H[s_2]\} &= H[s_1 s_2^\prime],\label{eq:b2}\\
\{ H[s_1], H[s_2]\} &= H_x[q^{xx}(s_1 s_2^\prime - s_1^\prime
  s_2)],\label{eq:b3} 
\end{align}
where $q^{xx}$ is a structure function.
Note that we have dropped the tilde which often appears on the top of the
structure function to distinguish it from the classical function.

The time evolution of the phase-space variables is given by

\begin{align}
\dot{E}^x       &= \{E^x,       H[N] + N_x[N^x] \},\label{eq:dynamic1}\\
\dot{K}_x       &= \{K_x,       H[N] + N_x[N^x] \},\label{eq:dynamic2}\\
\dot{E}^\varphi &= \{E^\varphi, H[N] + N_x[N^x] \},\label{eq:dynamic3}\\
\dot{K}_\varphi &= \{K_\varphi, H[N] + N_x[N^x] \},\label{eq:dynamic4}
\end{align}
where dots represent time derivatives.
The canonical conjugate momenta of the smearing functions $N$ and $N^x$ vanish
and thus they are nondynamical.

In addition, when evaluated on solutions at any given value of $t$ (on shell),
the diffeomorphism and Hamiltonian are constrained to vanish:

\begin{equation}
\mathcal{H}_x = 0 \quad\text{and}\quad \mathcal{H} = 0.
\label{eq:constraints}
\end{equation}
Both $\mathcal{H}_x$ and $\mathcal{H}$ are first-class constraints since the
Poisson brackets between them close. 
They thus generate gauge transformations on the phase space.

The dynamical flow on the phase space determines a spherically symmetric
geometry on a four-dimensional manifold $\mathcal{M}$.
The topology of this manifold is $\mathcal{M}^2\times \mathcal{S}^2$, where
$\mathcal{M}^2$ is a two-dimensional manifold and $\mathcal{S}^2$ a two-sphere.  
Together, the solution to the equations of
motion~\eqref{eq:dynamic1}-\eqref{eq:dynamic4} and
constraints~\eqref{eq:constraints} covariantly define the spherically symmetric 
metric on $\mathcal{M}$: 

\begin{equation}
ds^2 = -N^2 dt^2 + q_{xx} (dx + N^x dt)^2 + q_{\vartheta\vartheta} d\Omega^2,
\label{eq:metric}
\end{equation}
where $q_{ab}$ is the three-dimensional metric induced on the spatial
hypersurfaces foliating the manifold,
$q_{xx} = 1/q^{xx}$ and $q_{\vartheta\vartheta}$ is a scalar function on the
manifold $\mathcal{M}^2$,
$N$ and $N^x$ are the lapse function and shift vector field associated to the
observer frame with coordinate $t$, and
$d\Omega^2$ is the metric of the two-sphere on $\mathcal{S}^2$.
The coordinates $(t,x)$ correspond to any generic coordinates on
$\mathcal{M}^2$.

Due to the symmetry reduction of the scalar Hamiltonian to
spherical symmetry, the hypersurface deformed bracket does not lead to a
$q^{\vartheta\vartheta}$ component.
Thus the scalar $q_{\vartheta\vartheta}$ does not appear in the phase space and
its solution is not determined by the system.
Thus one may choose $q_{\vartheta\vartheta}$ to be any function of $E^x$, since
$E^x$ is the only phase-space variable that is a spacetime scalar.

When solving the equations of motion~\eqref{eq:dynamic1}-\eqref{eq:dynamic4} for a
given gauge, and thus picking a lapse $N$ and shift $N^x$, the form of the
metric~\eqref{eq:metric} in a certain coordinate system is obtained.
The gauge freedom on the phase space corresponds to diffeomorphism invariance
on the manifold~\cite{Alonso-Bardaji:2023vtl}. 
The algebra is referred to as the hypersurface deformation algebra.
The gauge transformations generated by $\mathcal{H}_x$ on the phase space
correspond to deformations on the hypersurface of constant $t$ on the manifold.
For the Hamiltonian constraint $\mathcal{H}$, gauge transformations generate
deformations normal to the hypersurfaces of constant $t$.

\section{Generalized uncertainty principle spacetime\label{sec:gup}}

In this section, we summarize for the unfamiliar reader the derivation of the
GUP spacetime~\cite{Fragomeno:2024tlh,Gingrich:2024mgk}.
The interior of a static spherically symmetric black hole expressed in
Ashtekar-Barbero variables~\cite{PhysRevD.36.1587} using Schwarzschild
coordinates is the Kantowski-Sachs~\cite{Ashtekar:2018cay} line element.
The components of the Ashtekar-Barbero connection and the densitized triad are
given by the configuration variables $b$ and $c$, and associated conjugate
momenta $p_b$ and $p_c$, respectively~\cite{Ashtekar:2005qt}.
The algebra of the canonical variables, is inherited from the algebra of the
Ashtekar-Barbero connection and the densitized triad, and is canonical.

Modifications of the Poisson algebra are made according to the GUP approach of
choosing a quadratic modification in the configuration variables.
The GUP correction is commonly written as an additional positive term in the
conjugate momentum squared.
Using the conjugate momentum in the GUP correction embodies a minimum length
scale.
Alternatively, using the conjugate variable embodies a minimum
momentum scale. 
Since we are interested in the GUP modifications of spacetime, we choose the
latter since the metric depends on the conjugate variables.
This choice no longer corresponds to a minimal uncertainty principle but
rather a GUP.
Guided by a prescription in loop quantum
gravity~\cite{Ashtekar:2006wn,Chiou:2008nm,Chiou:2012pg}
in which the quantum parameters of the models are made momentum
dependent\footnote{Without the momentum dependence, the
modified metric was found to not have the correct asymptotic limits in the metric
coefficients and the Kretschmann scalar.},
the GUP corrections are divided by the conjugate momentum squared giving 

\begin{equation}
\{b,p_b\} = \gamma \left( 1 + \frac{\beta_b b^2 L_0^4}{p_b^2} \right) \quad
\text{and} \quad
\{c,p_c\} = \gamma \left( 1 + \frac{\beta_c c^2 L_0^4}{p_c^2} \right),
\label{eq:poisson}
\end{equation}
where the definitions
\begin{equation}
Q_b = |\beta_b|\gamma^2 L_0^2 \quad \text{and} \quad
Q_c = |\beta_c|\gamma^2 L_0^6
\label{eq:parms}
\end{equation}
are used in the following.
The parameters $\beta_b$ and $\beta_c$ are dimensionless
negative\footnote{The negative signs are necessary to ensure
the reality condition of the metric over all radial coordinate values.}
constants, $\gamma$ is the Barbero-Immirzi parameter, and $L_0$ a
fiducial parameter introduced to make the symplectic structure, and hence the
definition of the Poisson brackets, well defined.
These intermediate variables are of no consequence here and their definitions can
be found in context in~\cite{Fragomeno:2024tlh}.
The distorted Poisson brackets~\eqref{eq:poisson} represent a minimal
modification that leads to consistent results regarding classical and
asymptotic limits, as well as singularity resolution. 

The equations of motion for the phase-space variables
typically lead to four coupled first-order differential equations in the time
and space coordinates. 
This is due to products of the phase-space variables in the Hamiltonian. 
For the Hamiltonian used here, it is common to pick the lapse function that
strategically decouples the phase-space variables. 
The lapse function so chosen causes the classical equations of motion for $b$
and $p_b$ to decouple from those of $c$ and $p_c$.
Solving the classical Hamiltonian equations of motion, and replacing the
solutions for $p_b$ and $p_c$ in the metric yields the interior metric.
The metric is then analytically extended to the full spacetime
by switching the timelike and radial spacelike coordinates $t \leftrightarrow
r$. 

The GUP diagonal line element in Schwarzschild coordinates
$(t,r,\vartheta,\varphi)$ can be written as~\cite{Gingrich:2024mgk} 

\begin{equation}
ds^2 = g_{tt}(r) dt^2 + g_{rr}(r) dr^2 + g_{\vartheta\vartheta}(r) d\Omega^2,
\label{eq:le1}
\end{equation}
where $\Omega^2 = d\vartheta^2 + \sin^2\vartheta d\varphi^2$ and the metric
coefficients are

\begin{align}
g_{tt}(r) &= -\left( 1 + \frac{Q_b}{r^2} \right) \left( 1 +
\frac{Q_c m^2}{r^8} \right)^{-1/4} \left[ 1 - \frac{2m}{r} \left( 1 +
\frac{Q_b}{r^2} \right)^{-1/2} \right],\\  
g_{rr}(r) &= \left( 1 + \frac{Q_c m^2}{r^8} \right)^{1/4}
\left[ 1 - \frac{2m}{r} \left( 1 + \frac{Q_b}{r} \right)^{-1/2}
\right]^{-1},\\  
g_{\vartheta\vartheta}(r) &= \left( 1 + \frac{Q_c m^2}{r^8} \right)^{1/4} r^2,  
\end{align}
where $m$ is the Arnowitt, Deser, Misner (ADM)
mass~\cite{PhysRev.122.997}, and $Q_b$ and $Q_c$ are GUP
parameters \eqref{eq:parms} of the model. 
We point out that the line element is not of the standard
$g_{rr} = -1/g_{tt}$ form of general relativity.
This is not uncommon for metrics derived from the Hamiltonian formalism such
as in loop quantum gravity.
The classical singularity in $g_{tt}$ is removed at $r=0$, but a new coordinate
singularity occurs in $g_{rr}$ at $r=0$.

When solving the Hamiltonian differential equations of motion
an integration constant is often need.
As we are free to choose the value of this constant, it not a fundamental
constant of the theory. 
It can often be identified by asymptotic or classical limits to correspond to a
mass parameter $m$.
However, the Hamiltonian constraint should not depend on it.
To avoid the mass $m$ appearing in the Hamiltonian, we absorb $m$ into the
GUP parameter $Q_c$ by defining $\bar{Q}_c = Q_c m^2$,
making $\bar{Q}_c$ a mass-dependent parameter.

For convenience, we define the following auxiliary functions of $r$:

\begin{equation}
  p_0(r) = 1 + \frac{Q_b}{r^2}, \quad
  p_1(r) = 1 + \frac{\bar{Q}_c}{r^8}, \quad\text{and}\quad
  p_2(r) = 1 - \frac{2m}{r} p_0^{-1/2}
\end{equation}
so that

\begin{equation}
g_{tt}(r) = -p_0 p_1^{-1/4} p_2, \quad
g_{rr}(r) =  p_1^{1/4} p_2^{-1}, \quad\text{and}\quad
g_{\vartheta\vartheta}(r) =  p_1^{1/4} r^2.
\end{equation}

The phase-space dynamics does not fix the area-radius function $r$.
In general relativity $r = \sqrt{x^2}$ and, for
simplicity~\cite{Alonso-Bardaji:2025hda}, this is also taken throughout.
To apply the method in~\cite{Alonso-Bardaji:2025hda}, we must make the metric
component $g_{\vartheta\vartheta}$ of the form $r^2$ by making a change of
variable  

\begin{equation}
r = \left( 1 + \frac{\bar{Q}_c}{r^{8}} \right)^{-1/8} \bar{r}.
\end{equation}
The transformed line element is
\begin{equation}
ds^2 = g_{tt}(\bar{r}) dt^2 + g_{rr}(\bar{r}) d\bar{r}^2 + \bar{r}^2 d\Omega^2,
\label{eq:le2}
\end{equation}
where
\begin{equation}
dr = \left( 1 + \frac{\bar{Q}_c}{r^8} \right)^{7/16} d\bar{r}
\end{equation}
has been used and the metric components become
\begin{alignat}{2}
g_{tt}(\bar{r}) &= -\left( 1 + \frac{Q_b}{r^2} \right) \left( 1 +
\frac{\bar{Q}_c}{r^8} \right)^{-1/4} \left[ 1 - \frac{2m}{r} \left(1 +
  \frac{Q_b}{r^2}\right)^{-1/2} \right] &&= -p_0 p_1^{-1/4} p_2,\\ 
g_{rr}(\bar{r}) &= \left( 1 + \frac{\bar{Q}_c}{r^8} \right)^{9/8} \left[1 -
\frac{2m}{r} \left( 1 + \frac{Q_b}{r^2}\right)^{-1/2} \right]^{-1}
&&= p_1^{9/8} p_2^{-1},\\
g_{\vartheta\vartheta}(\bar{r}) &= \bar{r}^2,
\end{alignat}
where $r$ must be re-expressed in terms of $\bar{r}$ in the above equations.
In making this coordinate transformation, we require $\bar{r} > (Q_c m^2)^{1/8}$
and note that the other metric components become complex for $\bar{r} <
(Q_b^4 - Q_c m^2)^{1/8}$. 
We should thus transform back to $r$ after deriving any Hamiltonian
dynamics. 

We will keep $p_0$, $p_1$, and $p_2$ in the $r$ variable and use the chain
rule to evaluate the derivatives with respect to $\bar{r}$.
We thus need

\begin{equation}
\frac{dp_0}{dr} = \frac{2(1-p_0)}{r},\quad
\frac{dp_1}{dr} = \frac{8(1-p_1)}{r},\quad\text{and}\quad
\frac{dp_2}{dr} = \frac{1-p_2}{p_0 r}.
\end{equation}

To apply the method~\cite{Alonso-Bardaji:2025hda}, we must cast the line element
into the form

\begin{equation}
ds^2 = -\left( h_1(\bar{r}) - \frac{2m}{h_2(\bar{r})} \right) dt^2 +
\frac{1}{h_3(\bar{r})} \left( h_1(\bar{r}) - \frac{2m}{h_2(\bar{r})}
\right)^{-1} d\bar{r}^2 + \bar{r}^2 d\Omega^2 \label{eq:deform} 
\end{equation}
with shape functions $h_1(\bar{r})$, $h_2(\bar{r})$, and $h_3(\bar{r})$.
None of the shape functions should depend on $m$ the mass parameter of
the Schwarzschild reference geometry. 
This form for the Schwarzschild-like geometry is
sufficiently general but not unique.
However to build on the work in~\cite{Alonso-Bardaji:2025hda} we take the same
form.

In the GUP case,

\begin{alignat}{2}
h_1(\bar{r}) &= \left( 1 + \frac{Q_b}{r^2} \right) \left( 1 +
\frac{\bar{Q}_c}{r^8} \right)^{-1/4} &&= p_0 p_1^{-1/4},\\
h_2(\bar{r}) &= \left( 1 + \frac{Q_b}{r^2} \right)^{-1/2}\left( 1 +
\frac{\bar{Q}_c}{r^8} \right)^{1/4} \bar{r} &&= p_0^{-1/2} p_1^{1/4} r,\\  
h_3(\bar{r}) &= \left( 1 + \frac{Q_b}{r^2} \right)^{-1} \left( 1 +
\frac{\bar{Q}_c}{r^8} \right)^{-7/8} &&= p_0^{-1} p_1^{-7/8}. 
\end{alignat}
where again $r$ must be re-expressed in terms of $\bar{r}$ in the above
equations. 

\section{Generalized uncertainty principle Hamiltonian\label{sec:hamiltonian}}

In this section, we write down the GUP Hamiltonian and derive the equations of
motion on the phase space.
We assume the diffeomorphism constraint remains unmodified by GUP corrections
to the classical theory.
The general Hamiltonian constraint can be written
as~\cite{Alonso-Bardaji:2025hda}

\begin{align}
\mathcal{H}_{(h_1,h_2,h_2)} &= \left[ -\frac{E^\varphi}{2h_2}
\frac{d(h_1 h_2)}{d\bar{r}} - \frac{K_\varphi^2 E^\varphi}{2h_2} \frac{d(h_2
h_3)}{d\bar{r}} + \frac{h_2^2}{2\sqrt{E^x}} \left(
\frac{(E^x)^{\prime\prime}}{E^\varphi} - \frac{(E^x)^\prime
  (E^\varphi)^\prime}{(E^\varphi)^2} \right)\right.\nonumber\\ 
  &\quad \left.\left. - \left( \frac{2h_2}{\sqrt{E^x}} - 3\frac{dh_2}{d\bar{r}}
\right) \frac{h_2}{E^x} \frac{[(E^x)^\prime]^2}{8E^\varphi} -2\sqrt{E^x} h_3
K_x K_\varphi \right] \right|_{\bar{r}=\sqrt{E^x}}.
\label{eq:hamiltonian}
\end{align}
This expression is derived by starting with a general
Hamiltonian with six free functions of $E^x$.
For simplicity, $\bar{r} = \sqrt{E^x}$ is chosen.
Given the form~\eqref{eq:deform} the free functions are solved in terms of the
shape functions and $\bar{r}$ to give~\eqref{eq:hamiltonian}.

The dynamical flow on the phase space given by the Hamiltonian

\begin{equation}
H = \int (N\mathcal{H}_{(h_1,h_2,h_3)} + N^x \mathcal{H}_x) dx
\end{equation}
covariantly defines the deformed Schwarzschild geometry~\eqref{eq:deform} for
any choice of the shape functions.
The solution of the Hamiltonian constraints in any gauge will give the same
geometry since the construction is covariant.
The Hamiltonian is unique up to canonical transformations that do not include
derivative terms and leave invariant the diffeomorphism constraint.
The expression~\eqref{eq:hamiltonian} can be considered as a family of Hamiltonian
constraints. 
The GUP Hamiltonian constraint is one member of the family.

Substitution of the GUP shape functions into \eqref{eq:hamiltonian} gives the
GUP Hamiltonian constraint 

\begin{equation}
\begin{aligned}
\mathcal{H} &= -\frac{E^\varphi}{2\sqrt{E^x}} 
p_1^{5/16}
- \frac{E^\varphi}{2\sqrt{E^x}} K_\varphi^2
p_0^{-2} p_1^{-21/16} \left[ p_0\left(-5+9p_1\right) - 3p_1 \right]\\
&\quad + \frac{\sqrt{E^x}}{2} \left( \frac{(E^x)^{\prime\prime}}{E^\varphi}
- \frac{(E^x)^\prime (E^\varphi)^\prime}{(E^\varphi)^2} \right) p_0^{-1}
p_1^{1/4}\\
  &\quad + \frac{\sqrt{E^x}[(E^x)^\prime]^2}{8E^x E^\varphi} p_0^{-2}
p_1^{-3/16} \left[ -2p_0\left(-3+p_1^{7/16}\right) - 3p_1 \right]\\   
&\quad -2 \sqrt{E^x}K_x K_\varphi p_0^{-1} p_1^{-7/8}.\label{eq:ham}
\end{aligned}
\end{equation}
The Hamiltonian constraint has been written so that each of the vacuum general
relativity terms are multiply by GUP corrections.
In the limit $Q_b\to 0$ and $Q_c\to 0$, we recover the Hamiltonian constraint
of vacuum spherically symmetric general relativity.

The Hamiltonian cannot be understood as one corresponding to vacuum general
relativity plus certain additive corrections.
Since the GUP parameters do not appear as additive
contribution to the Hamiltonian constraint, they can not be understood as being
originated from a minimal coupling to matter fields.

The general structure function is expressed in terms of the
six free functions in~\cite{Alonso-Bardaji:2025hda} by computing the
hypersurface deformed bracket for the scalar Hamiltonian constraint.
Then expressing the free functions again in terms of the shape functions, and
ultimately the GUP auxiliary functions, the structure function is 

\begin{equation}
  q^{xx} = \frac{h_2^2 h_3}{(E^\varphi)^2}
  = \frac{E^x}{(E^\varphi)^2} p_0^{-2} p_1^{-5/8}.\label{eq:sf}
\end{equation}

The hypersurface deformation brackets~\eqref{eq:b1}-\eqref{eq:b3} should be
satisfied by construction.
The diffiomorphism bracket~\eqref{eq:b1} is identically satisfied since the
classical diffiomorphism constraint has not changed.
As a check on the Hamiltonian constraint~\eqref{eq:ham} and structure
function~\eqref{eq:sf}, we have explicitly calculated the
brackets~\eqref{eq:b2}-\eqref{eq:b3} and shown the algebra to close.
(See Appendix~\ref{sec:appB} for some details.)

\subsection{Equations of motion}

Using the dynamic-flow equations~\eqref{eq:dynamic1}-\eqref{eq:dynamic4}, the GUP
equations of motion are 

\begin{align}
\dot{E}^x &= 2 N \sqrt{E^x} K_\varphi p_0^{-1} p_1^{-7/8} + N^x
(E^x)^\prime,\label{eq:eom1}\\ 
\dot{E}^\varphi &= \frac{N}{\sqrt{E^x}} \left[ K_\varphi E^\varphi p_0^{-2}
p_1^{-21/16} \left[ p_0(-5+9p_1) -3p_1 \right] +
2K_x E^x p_0^{-1} p_1^{-7/8}\right] + (N^x E^\varphi)^\prime,\label{eq:eom2}\\
\dot{K}_\varphi &= - \frac{N}{2\sqrt{E^x}} \left[ p_1^{5/16} +
  K_\varphi^2p_0^{-2} p_1^{-21/16} \left[ p_0(-5+9p_1)-3p_1
    \right]\right]\nonumber\\
  &\quad -\frac{N}{2\sqrt{E^x}} \left( \frac{(E^x)^\prime}{2E^\varphi}
\right)^2p_0^{-2} p_1^{-3/16} \left[ -2p_0\left(-3+p_1^{7/16}\right) - 3p_1
  \right]  + \left(N\sqrt{E^x}p_0^{-1} p_1^{1/4}\right)^\prime
\frac{(E^x)^\prime}{2(E^\varphi)^2}\nonumber\\ 
&\quad + N^x K_\varphi^\prime.\label{eq:eom3}  
\end{align}
These equations are written similarly to a typical form of writing the classical
equations to allow one to easily view the multiplicative GUP corrections.
The equation for $\dot{K}_x$ has not been written as it is in excess of 50
terms and unlikely to be insightful.
Setting $p_1 = p_2 \to 1$, gives the classical results.

\section{Choice of gauge\label{sec:gauge}}

In this section, we determine the phase-space variables, lapse, shift, and
thus line element, in different gauges.
The construction of the metric in different charts has already been obtained by
coordinate transformations~\cite{Fragomeno:2024tlh,Gingrich:2024mgk}, but the
aim here is not to see the different forms of the metrics themselves but to
validate the covariance. 
Does the theory starting from a covariant GUP Hamiltonian and using the canonical
formalism obtain the same results as a non-covariant formalism using GUP
distorted Poisson brackets? 

Since the constrain equations and equations of motion can sometimes be
difficult to solve, we can be guided by the coefficients in the general line
element. 
The spherically symmetric metric in the ADM formalism~\cite{PhysRev.117.1595}
can be expanded as  

\begin{equation}
\begin{aligned}
ds^2 &= -N^2 dt^2 + \frac{1}{q^{xx}} \left(dx + N^x dt\right)^2 + E^x
d\Omega^2\\ 
&= -\left[N^2 - \frac{(N^x)^2}{q^{xx}} \right] dt^2 +
\frac{1}{q^{xx}} dx^2 + 2 \frac{N^x}{q^{xx}} dx dt + E^x d\Omega^2.
\end{aligned}
\end{equation}
Written in this form, the line element allows us to identify the correspondence
to the GUP metric and thus aid in obtaining the line element for a particular
gauge. 

\subsection{Static gauge}

To follow~\cite{Alonso-Bardaji:2025hda}, we must take $x = \sqrt{E^x}$. 
Applying this partial gauge, the diffeomorphism constraint gives 

\begin{equation}
K_x = \frac{E^\varphi}{2\bar{r}} K_\varphi^\prime.
\label{eq:sdiff}
\end{equation}
Conservation of this gauge condition requires $\dot{E}^x = 0$, allowing the
equation of motion~\eqref{eq:eom1} to be written as

\begin{equation}
  N^x = - N K_\varphi p_0^{-1} p_1^{-7/8}.
  \label{eq:seom}
\end{equation}

\subsubsection{Schwarzschild gauge}

In the Schwarzschild gauge, $N^x = 0$ which giving $K_\varphi = 0$ by
\eqref{eq:seom} and thus $K_x = 0$ by \eqref{eq:sdiff}.

It remains to determine $E^\varphi$ and $N$.
The Hamiltonian constraint and remaining equation of motion become

\begin{align}
\mathcal{H} = 0 &= -\frac{E^\varphi}{2x} p_1^{5/16}
+ \frac{3x}{2E^\varphi} p_0^{-2} p_1^{-3/16} (2p_0-p_1)
- \frac{x^2(E^\varphi)^\prime}{(E^\varphi)^2} p_0^{-1} p_1^{1/4}\\
\dot{K}_\varphi &= -\frac{N}{2x} p_1^{5/16}
-\frac{Nx}{2(E^\varphi)^2} p_0^{-2} p_1^{-3/16} \left( 2p_0-p_1 \right) +
\frac{N^\prime x^2}{(E^\varphi)^2} p_0^{-1} p_1^{1/4}.   
\end{align}
The structure function is 

\begin{equation}
  q^{xx} = \frac{x^2}{(E^\varphi)^2} p_0^{-2} p_1^{-5/8}.
\end{equation}

The Hamiltonian constraint can be solve for $E^\varphi$.
Given $E^\varphi$ and the static requirement of $\dot{K}_\varphi = 0$, we can
solve for $N$. 
These solutions, along with their derivatives are

\begin{align}
E^\varphi &= p_0^{-1} p_1^{1/4} p_2^{-1/2} x =
\left(1+\frac{Q_b}{r^2}\right)^{-1} \left(1+\frac{Q_bm^2}{r^8}\right)^{1/4}
\left[1-\frac{2m}{r}\left(1+\frac{Q_b}{r^2}\right)^{-1/2}\right]^{-1/2},\\   
(E^\varphi)^\prime &= -\frac{1}{2}p_0^{-2}p_1^{-3/16} \left[
p_1p_2^{-3/2} - 3(2p_0-p_1)p_2^{-1/2} \right],\\
N &= p_0^{1/2} p_1^{-1/8} p_2^{1/2} = \left(1+\frac{Q_b}{r^2}\right)^{1/2}
\left(1+\frac{Q_cm^2}{r^8}\right)^{-1/8}
\left[1+\frac{2m}{r}\left(1+\frac{Q_b}{r^2}\right)^{-1/2}\right]^{1/2},\\  
N^\prime &= \frac{1}{2} p_0^{-1/2} p_1^{-9/16} \left[p_1p_2^{-1/2}
-(2p_0-p_1)p_2^{1/2}\right]x^{-1}. 
\end{align}
The structure function becomes

\begin{equation}
  q^{xx} = p_1^{-9/8} p_2
  = \left( 1 + \frac{Q_cm^2}{r^8} \right)^{-9/8} \left[ 1 - \frac{2m}{r} \left(
    1 + \frac{Q_b}{r^2} \right)^{-1/2}\right].
\end{equation}
The line element obtained in the Schwarzschild gauge corresponds the original
line element~\eqref{eq:le2} obtained using deformed Poisson brackets.
After writing down the resulting line element, one should transform $x =
\bar{r}$ back to $r$ to obtained the untransformed line element~\eqref{eq:le1}
in the Schwarzschild coordinate $r$.
\subsubsection{Gullstrand-Painlev{\'e} gauge}

For a $g_{tt} \neq -1/g_{rr}$ non-symmetric metric, the spatial part of the
metric in the Gullstrand-Painlev{\'e} gauge is $-g_{tt} g_{rr}$, rather
than flat. 
With $E^x = x^2$, the spatial part of the metric is

\begin{equation}
-g_{tt} g_{rr} = \frac{1}{q^{xx}} \quad\Rightarrow\quad E^\varphi = p_0^{-1/2} 
p_1^{1/8} x = \pm\left(1+\frac{Q_b}{r^2} \right)^{1/2} \left(
1+\frac{Q_bm^2}{r^8} \right)^{1/8} x. 
\end{equation}

The diffeomorphism constraint gives

\begin{equation}
K_x = \frac{K_\varphi^\prime}{2} p_0^{1/2} p_1^{1/8},\label{eq:diffPG}
\end{equation}
which allows us to determine $K_x$ once $K_\varphi$ is known.

Conservation of the gauge conditions for $K_x$ and $K_\varphi$ gives

\begin{align}
  \dot{E}^x = 0 &= 2x(N K_\varphi p_0^{-1}p_1^{-7/8} + N^x),\\
  \dot{E}^\varphi = 0 &= N\left(K_\varphi + 2xK_x p_0^{-1} p_1^{-7/8}\right)
  + \left(xN^x p_0^{-1/2}p_1^{1/8}\right)^\prime.
\end{align}

Direct inspection of the line element~\cite{Gingrich:2024mgk} gives

\begin{equation}
  \sqrt{(-g_{tt}g_{rr})(1+g_{tt})} = \frac{N^x}{q^{xx}}
  \quad\Rightarrow\quad 
  N_x = \sqrt{\frac{1-p_0p_1^{-1/4}p_2}{p_0p_1^{7/8}}}
\end{equation}
and

\begin{equation}
  g_{tt} = -N^2 + \frac{(N^x)^2}{q^{xx}} \quad\Rightarrow\quad N = 1.
\end{equation}

Using $\dot{E}^x = 0$, gives

\begin{equation}
K_\varphi = -\left( 1 - p_0 p_1^{-1/4} p_2 \right)^{1/2} p_0^{1/2} p_1^{7/16}.
\end{equation}
Thus $K_x$ can be determined by differentiating $K_\varphi$ and substituting
into~\eqref{eq:diffPG}. 
After transform $x = \bar{r}$ back to $r$ we obtain the Gullstrand-Painl{\'e}ve
line element in~\cite{Gingrich:2024mgk}.
\subsection{Homogeneous gauge}

In the homogeneous gauge, the function $E^x$ on the phase space obeys
$(E^x)^\prime = 0$ and $\dot{E}^x \ne 0$, in general.
To complete the gauge fixing we require $(E^\varphi)^\prime = 0$ and $N^x = 0$.
This in turn gives $K_\varphi^\prime  = 0$, $K_x^\prime = 0$, and $N^\prime =
0$ on the constraint surface~\cite{Alonso-Bardaji:2022ear}.

The diffeomorphism constraint identically vanishes.
The Hamiltonian constraint reduces to

\begin{equation}
\mathcal{H} = -\frac{E^\varphi}{2\sqrt{E^x}} p_0^{-2} p_1^{-21/16} \left[
p_0^{2} p_1^{13/8} + K_\varphi^2 \left(p_0(-5+9p_1)-3p_1\right) \right] -
2\sqrt{E^x} K_x K_\varphi p_0^{-1} p_1^{-7/8}. 
\end{equation}

Solving $\mathcal{H} = 0$ for $K_x$ gives

\begin{equation}
K_x = -\frac{E^\varphi}{4E^x K_\varphi} p_0^{-1} p_1^{-7/16} \left[ p_0^2
  p_1^{13/8} +  K_\varphi^2 \left[ p_0(-5+9p_1)-3p_1\right]\right].
\label{eq:static}
\end{equation}

Recalling that the time coordinate is spacelike and the radial
coordinate is timelike in the homogeneous gauge, inspection of the line element
gives 



\begin{align}
N &= p_1^{9/16} (-p_2)^{-1/2},\\
E^\varphi &= \sqrt{E^x} p_0^{-1/2} p_1^{-7/16} (-p_2)^{1/2}.
\end{align}
Since $E^x$, $p_0$, and $p_1$ are positive, one could choose
to absorb the negative sign into the definition of $p_2$ to give the usual form

\begin{equation}
-p_2 = \frac{2m}{\sqrt{E^x}}p_0^{-1/2} - 1.
\end{equation}
in which the coefficients are written in the Kantowski-Sachs line element.
Since $\sqrt{E^x}$ should play the role of time, its equation of motion should
give

\begin{equation}
\dot{E}^x = 2 \sqrt{E^x}.
\end{equation}
or

\begin{align}
\dot{E}^x &= 2N\sqrt{E^x} K_\varphi p_0^{-1} p_1^{-7/8}\nonumber\\
&= 2\sqrt{E^x} K_\varphi p_0^{-1} p_1^{-5/16} p_2^{-1/2} \quad\Rightarrow\quad
K_\varphi = p_0 p_1^{5/16} p_2^{1/2}.
\end{align}
Substitution into~\eqref{eq:static} give $K_x$.

When comparing the results here with the homogeneous results
of~\cite{Fragomeno:2024tlh,Gingrich:2024mgk} there is no direct connection
between the phase-space variables or the Hamiltonian constraint.
However, the identification $\tilde{t} = \sqrt{E^x}$ gives a line element
identical to the Kantowski-Sachs line element with the same lapse and structure
functions.  
This result demonstrates the covariance of the Hamiltonian constraint presented
here~\eqref{eq:ham} and gives an identical line element to that derived by
the distorted Poisson bracket motivated by the GUP.

\section{Matter coupling\label{sec:matter}}

Coupling the GUP Hamiltonian to matter is essential for studying its dynamical
behaviour.
While expressions for Hamiltonians for various fields are known, it seems here
would be a natural place to calculate them in GUP spacetime allowing them to be
used in future work.

One typically starts with scalar fields and dust as they are easiest and
elucidate basic dynamical properties.
To couple to scalar matter, we introduce an additional pair of conjugate
variables 

\begin{equation}
\{ \phi(t,x_1),P_\phi(t,x_2)\} = \delta(x_1-x_2).
\end{equation}

The total Hamiltonian becomes

\begin{equation}
H_\text{total} = \int \left[ (\mathcal{H}_x + \mathcal{H}_x^\mathrm{m}) N^x +
  (\mathcal{H} + \mathcal{H}^\mathrm{m}) N\right] dx,
\end{equation}
where $\mathcal{H}_x$ and $\mathcal{H}$ are the gravitational diffeomorphism
and Hamiltonian constrains respectively, while $\mathcal{H}_x^\mathrm{m}$ and
$\mathcal{H}^\mathrm{m}$ 
correspond to the matter contributions.
The scalar diffeomorphism constraint is

\begin{equation}
\mathcal{H}_x^\mathrm{m} = \phi^\prime P_\phi
\end{equation}
and the Hamiltonian constraint is (see Appendix~\ref{sec:appC})

\begin{align}
  \mathcal{H}^\mathrm{m} &= \frac{1}{2} \left[\ \frac{P_\phi^2}{\sqrt{q}}
  + \sqrt{q} q^{ab} \partial_a\phi \partial_b\phi + \sqrt{q} V(\phi)
      \right]\nonumber\\  
  &= \frac{ p_0^{-1} p_1^{-5/16} \sqrt{E^x}} {2E^\varphi} \left[
\frac{P_\phi^2}{E^x \sin\vartheta} + E^x \sin\vartheta (\phi^\prime)^2 \right] +
  \frac{p_0p_1^{5/16}\sqrt{E^x} E^\varphi \sin\vartheta}{2} V(\phi).
\end{align}
where
\begin{equation}
P_\phi = \frac{\sqrt{q}}{N} \left( \dot{\phi} - N^a\partial_a\phi\right)
= \frac{p_0 p_1^{5/16} \sqrt{E^x}E^\varphi \sin\vartheta}{N} \left( \dot{\phi}
- N^a\partial_a\phi\right). 
\end{equation}

The Klein-Gordon equation on the curved covariant metric has been shown to have
anomaly-free constraint brackets and to respect covariance of both the
spacetime and the matter field in spherical symmetry~\cite{Bojowald:2024lhr}. 
The gravitational Hamiltonian constraint does not depend on $\phi$ or $P_\phi$
as it is the vacuum background, while the scalar Hamiltonian constraint depends
only quadratically on $\phi$ and $P_\phi$.
It can be shown~\cite{Bojowald:2024lhr} that under these circumstances the
back reaction can be neglected, and thus will not be effected by perturbative
dynamics. 

Likewise for dust, the diffeomorphism constraint is

\begin{equation}
\mathcal{H}_x^\mathrm{m} = T^\prime P_T
\end{equation}
and the Hamiltonian constraint is~\cite{Brown:1994py} (see
Appendix~\ref{sec:appC})

\begin{equation}
\mathcal{H}^\mathrm{m} = P_T \sqrt{q^{ab} \partial_a T
  \partial_b T + 1} = P_T \sqrt{p_0^{-2} p_1^{-5/8}
  \frac{E^x}{(E^\varphi)^2}(T^\prime)^2 + 1}.  
\end{equation}
where
\begin{equation}
P_T = \frac{\sqrt{q}M}{N} \left( \dot{T} - N^a\partial_a T\right)
= \frac{p_0 p_1^{5/16} \sqrt{E^x}E^\varphi \sin\vartheta M}{N} \left( \dot{T} -
N^a\partial_a T\right). 
\end{equation}

\section{Summary\label{sec:discussion}}

Starting from a metric obtained using deformed Poisson brackets, we have
derived a Hamiltonian that allows a canonical and covariant formalism to be
defined.
The Hamiltonian is quadratic in the derivatives and second-order in the first
derivatives.
Since the formalism used to determine the Hamiltonian involves several
functions that need to be determined, the Hamiltonian constraint may not be
unique.
The dynamical flow covariantly defines a static and spherical symmetric
four-dimensional geometry corresponding to the GUP metric.

The lapse, shift, and phase-space variables have been calculated in the
Schwarzschild and Gullstrand-Painl{\'e}ve gauges and shown to lead to the
correct line elements.
The homogeneous gauge has also been determined, and with a proper
identification of the time variable reproduces the metric from which the GUP was
derived.
Thus, three choices of gauge have been shown to give charts that would be
obtained from different choices of coordinates.

Simple scalar and dust matter fields have been coupled to the geometry.
The procedure used here could help elucidate the dynamical origin of some
static geometries.
The coupling to a scalar field would allow the study of quasinormal modes or
Hawking evaporation, to name just two examples.
Coupling to dust allows gravitational collapse to be studied, as well as
providing a clock.

Although starting from heuristic arguments, the GUP spacetime, which can give
black hole, wormhole, and remnant solutions, has been cast into a consistent
canonical formalism that can covariantly give different charts depending on
the choice of gauge.
The next step is to study the dynamics.

\appendix 
\section{Deformed Poisson brackets in the static gauge\label{sec:appA}} 

In this appendix, we show that it is not possible to use the same
procedure as in~\cite{Fragomeno:2024tlh,Gingrich:2024mgk} to determine GUP
corrections using deformed Poisson brackets in the static gauge.
The GUP line element was derived using the interior of a static spherically
symmetric black hole expressed in Ashtekar-Barbero
variables~\cite{PhysRevD.36.1587} using Schwarzschild coordinates.
This is a cosmological based mechanical mini-superspace model with no $r$
dependence.
It is not possible impose a static gauge choice.

To attempt to circumvent this limitation, let's apply the same deformed
Poisson bracket procedure in~\cite{Fragomeno:2024tlh,Gingrich:2024mgk} to a
field theory midi-superspace model with $r$ dependence.
We need only consider one of the conjugate phase-space pair of fields
$(K_x,E^x)$, say. 
the canonical Poisson bracket is

\begin{equation}
\left\{ K_x(t,x_1), E^x(t,x_2) \right\}_\mathrm{canonical} = \delta(x_1-x_2).
\end{equation}
The deformed Poisson bracket can be specified by a function $F$ which in
general could be a function of the pair of conjugate phase-space fields and a
constant $\beta$.
The deformed Poisson bracket can be written as

\begin{equation}
\left\{ K_x(t,x_1), E^x(t,x_2) \right\}_\mathrm{deformed} =
\left\{ K_x(t,x_1), E^x(t,x_2) \right\}_\mathrm{canonical} [ 1 +
F(K_x,E^x,\beta)],
\end{equation}
The equation of motion for $E^x$, say, is

\begin{equation}
\dot{E^x} = \left\{ E^x,H \right\}_\text{deformed}
 = \left\{ E^x,H \right\}_\text{canonical} [1 + F(K_x,E^x,\beta)].
\end{equation}
In the static gauge,

\begin{equation}
\dot{E^x} = 0 \quad \Rightarrow \quad \left\{ E^x,H \right\}_\text{canonical} =
0. 
\end{equation}
The equation of motion is canonical and GUP corrections have no effect in
static gauge.
A theory based on multiplicative Poisson bracket deformations is thus not
canonical nor covariant.

\section{Hypersurface deformation algebra\label{sec:appB}}

In this appendix we discuss the calculation of the hypersurface deformation
algebra which is rarely discussed, beyond stating that it's a long but
straightforward calculation.
The hypersurface deformation brackets~\eqref{eq:b1}-\eqref{eq:b3} should be
satisfied by construction.
The diffiomorphism bracket~\eqref{eq:b1} is identically satisfied since the
classical diffiomorphism constraint has not been changed.

Each term in the classical Hamiltonian constraint contains a function of
$\sqrt{E^x}$.
The GUP modifications appear as multiplicative factors $p_0$ and $p_1$ to
various powers, which are also functions of $\sqrt{E^x}$, and thus just change
the function in each term of the classical Hamiltonian constraint.
When calculating the mixed hypersurface bracket~\eqref{eq:b2} the
differentials of the functions of $\sqrt{E^x}$ are eliminated by integrating by
parts thus leaving the Hamiltonian constraint effectively unaltered.
The same can be done with the GUP modified Hamiltonian and the hypersurface
bracket~\eqref{eq:b2} is satisfied.

The bracket~\eqref{eq:b3} requires the most work and is less intuitive.
The terms in the Hamiltonian constraint can be numbered as $A_1, A_2, A_3,
\ldots A_6$. 

\begin{equation}
\mathcal{H} = A_1 + A_2 + A_3 + A_4 + A_5 + A_6,
\end{equation}
where

\begin{align*}
A_1 &= -\frac{E^\varphi}{2\sqrt{E^x}} p_1^{5/16},\\
A_2 &= -\frac{E^\varphi}{2\sqrt{E^x}} K_\varphi^2
       p_0^{-2} p_1^{-21/16} \left[ p_0\left(-5+9p_1\right) - 3p_1 \right],\\
A_3 &= \frac{\sqrt{E^x}}{2} \frac{(E^x)^{\prime\prime}}{E^\varphi} p_0^{-1}
p_1^{1/4},\\ 
A_4 &= - \frac{\sqrt{E^x}}{2} \frac{(E^x)^\prime (E^\varphi)^\prime}{(E^\varphi)^2}
p_0^{-1} p_1^{1/4},\\
A_5 &= \frac{\sqrt{E^x}[(E^x)^\prime]^2}{8E^x E^\varphi} p_0^{-2}
p_1^{-3/16} \left[ -2p_0\left(-3+p_1^{7/16}\right) - 3p_1 \right],\\   
A_6 &= -2 \sqrt{E^x}K_x K_\varphi p_0^{-1} p_1^{-7/8}.
\end{align*}
The only nonzero brackets are between terms involving $K$ and the corresponding
derivatives of $E$, that is, $\{A_2,A_4\}$, $\{A_6,A_5\}$, $\{A_6,A_4\}$,
and $\{A_6,A_3\}$.
In the classical case, bracket $\{A_2,A_4\}$ cancels bracket $\{A_6,A_5\}$.
The bracket $\{A_6,A_4\}$ gives two nonzero terms: the $K_x$ term is a desired
term but the $K_\varphi$ term is undesirable and must be cancelled.
The bracket $\{A_6,A_3\}$ consists of the product of two total derivatives
which can strategically be integrated by parts to give another one of the desired
terms and a second term which cancels the previously undesired term.
The GUP case is even less intuitive.
In this case, $\{A_2,A_4\}$ and $\{A_6,A_5\}$ do not cancel, but together
generate three terms, none of which are desirable.
Similar to the classical case, the bracket $\{A_6,A_4\}$ gives two nonzero
terms: the $K_x$ term is a desired term but the $K_\varphi$ term is undesirable
and must be cancelled.
We now have four undesirable terms.
The bracket $\{A_6,A_3\}$ again gives the product of two total derivatives
but this time a strategic integration by parts gives five terms; one
desired term and four terms that cancel the previous undesirable terms.
The extra terms in the integration by parts come from the differentiation of
the $p_0, p_1$ factors.  

\section{Matter Hamiltonians\label{sec:appC}}

\subsection{Scalar Hamiltonian}

In this appendix, we derive the equation of motion for a scalar field using the
canonical formalism.
The Lagrangian density for a scalar field is

\begin{equation}
\mathcal{L}_\mathrm{m} = -\frac{1}{2} \sqrt{-g} \left[ g^{\mu\nu} \partial
\phi_\mu \partial_\nu \phi + V(\phi) \right].
\end{equation}
To evaluate this in curved spacetime, we need the inverse metric.
First, the line element in ADM form can be written as 

\begin{equation}
ds^2 = -N^2 dt^2 + q_{ab} (dx^a + N^a dt)(dx^b + N^b dt)
\end{equation}
with non-zero metric components

\begin{equation}
  g_{00} = -N^2 + q_{ab} N^a N^b, \quad
  g_{0a} = g_{a0} = q_{ab} N^a, \quad \text{and} \quad
    g_{ab} = q_{ab}.
\end{equation}
The non-zero inverse metric components are

\begin{equation}
  g^{00} = -\frac{1}{N^2}, \quad
  g^{0a} = g^{a0} = \frac{N^b}{N^2}, \quad \text{and} \quad
    g^{ab} = q^{ab} - \frac{N^aN^b}{N^2}.
\end{equation}

We will also need the determinant of the metric $g$ which can be expressed as 

\begin{equation}
\sqrt{-g} = N \sqrt{q}.
\end{equation}

The Lagrangian density becomes

\begin{equation}
\mathcal{L}_\mathrm{m} = -\frac{1}{2} N\sqrt{q} \left[ -\frac{1}{N^2}
  (\dot{\phi})^2 + 2\frac{N^a}{N^2} \dot{\phi} \partial_a \phi + \left( q^{ab}
  -\frac{N^aN^b}{N^2} \right) \partial_a\phi \partial_b\phi + V(\phi) \right].
\end{equation}

The conjugate momentum of the scalar field is

\begin{equation}
P_\phi = \frac{\delta \mathcal{L}_\mathrm{m}}{\delta\dot{\phi}} =
\frac{\sqrt{q}}{N} (\dot{\phi} - N^a \partial_a \phi).
\end{equation}
To eliminate the time derivatives from the Lagrangian, we perform the following
manipulations.
Inverting the momentum equation gives

\begin{equation}
  \dot{\phi} = \frac{N}{\sqrt{q}} P_\phi + N^a \partial_a \phi.
\end{equation}
Squaring gives

\begin{equation}
\frac{\sqrt{q}}{N} (\dot{\phi})^2 = \frac{N}{\sqrt{q}} P_\phi^2 + 2 N^a P_\phi
\partial_a \phi + \frac{\sqrt{q}}{N} N^a N^b \partial_a \phi \partial_b
\phi.\label{eq:w1}
\end{equation} 
To enable the Legendre transformation, we multiply the momentum equation by
$\dot{\phi}$ to get

\begin{equation}
\frac{\sqrt{q}}{N} \left[ (\dot{\phi})^2 - N^a
    \dot{\phi} \partial_a \phi \right] = P_\phi \dot{\phi}.\label{eq:w2}
\end{equation} 
Subtracting one-half of \eqref{eq:w1} from \eqref{eq:w2} and substituting the
result into the first two terms of the Lagrangian density gives

\begin{equation}
\mathcal{L}_\mathrm{m} = P_\phi \dot{\phi} - N^a P_\phi \partial_a \phi -
  \frac{N}{2} \left( \frac{P_\phi^2}{\sqrt{q}} + \sqrt{q} q^{ab} \partial_a\phi  
\partial_b\phi + \sqrt{q}V  \right),
\end{equation}
where we can easily identify the diffeomorphism and Hamiltonian constraints.

Performing a Legendre transformation gives

\begin{equation}
P_\phi \dot{\phi} - \mathcal{L}_\mathrm{m} = N\mathcal{H}^\mathrm{m} + N^x
\mathcal{H}_x^\mathrm{m}
\end{equation}
and the constraints are

\begin{align}
  \mathcal{H}_x^\mathrm{m} &= P_\phi \partial_x \phi,\\
  \mathcal{H}^\mathrm{m} &= \frac{1}{2} \left( \frac{P_\phi^2}{\sqrt{q}} +
  \sqrt{q} q^{ab} \partial_a\phi \partial_b\phi + \sqrt{q} V\right).
\end{align}

For spherical symmetry, we use

\begin{equation}
\sqrt{q} = \sqrt{q_{xx}} q_{\vartheta\vartheta} \sin\vartheta
\end{equation}
to obtain

\begin{equation}
\dot{\phi} = \frac{NP_\phi}{\sqrt{q}} + N^x\partial_x\phi
= \frac{N\sqrt{q^{xx}}P_\phi}{q_{\vartheta\vartheta}\sin\vartheta} +
N^x\partial_x\phi.
\label{eq:phid}
\end{equation}
and

\begin{align}
\dot{P}_\phi &= \partial_b\left( N\sqrt{q} q^{ab} \partial_a \phi\right)
-\frac{N}{2} \sqrt{q} \frac{\partial V}{\partial \phi} + \partial_x (N^x
  P_\phi)\nonumber\\
&= \partial_x \left( N\sqrt{q^{xx}} q_{\varphi\varphi} \sin\vartheta \partial_x 
  \phi\right) + 
  N \sqrt{q_{xx}} \partial_\vartheta \left( 
  \sin\vartheta \partial_\vartheta \phi\right) +
  \frac{ N\sqrt{q_{xx}}  
  \partial_\varphi^2 \phi}{\sin\vartheta}\nonumber\\
  &\quad -\frac{N}{2} \sqrt{q_{xx}}
  q_{\vartheta\vartheta} \sin\vartheta \frac{\partial
    V}{\partial \phi} 
  + \partial_x (N^x 
  P_\phi)\nonumber\\
&= \sin\vartheta \left[ \partial_x \left( N\sqrt{q^{xx}} q_{\varphi\varphi}
    \partial_x \phi\right) + 
  N \sqrt{q_{xx}} \left( \frac{\partial_\vartheta \left( 
  \sin\vartheta \partial_\vartheta \phi\right)}{\sin\vartheta} +
  \frac{\partial_\varphi^2 \phi}{\sin^2\vartheta}\right) \right.\nonumber\\
  &\quad \left. -\frac{N}{2} \sqrt{q_{xx}}
  q_{\vartheta\vartheta} \frac{\partial
    V}{\partial \phi}\right] + \partial_x (N^x P_\phi)\nonumber\\
&= \sin\vartheta \left[ \partial_x \left( N\sqrt{q^{xx}} q_{\varphi\varphi}
    \partial_x \phi\right) + 
  N \sqrt{q_{xx}} \Delta^\varphi -\frac{N}{2} \sqrt{q_{xx}}
  q_{\vartheta\vartheta} \frac{\partial
    V}{\partial \phi} \right]+ \partial_x (N^x P_\phi),
  \label{eq:Pd}
\end{align}
where

\begin{equation}
\Delta^\varphi = \frac{1}{\sin\vartheta} \partial_\vartheta (\sin\vartheta
\partial_\vartheta )+ \frac{\partial_\varphi^2}{\sin^2\vartheta}. 
\end{equation}
is the Laplacian on a two-sphere.

Equations~\eqref{eq:phid} and \eqref{eq:Pd} are two first-order equations.
We can write them as a single second-order equation by a partial gauge fixing
of $N^x = 0$:

\begin{equation}
\ddot{\phi} = \frac{N \sqrt{q^{xx}}}{q_{\vartheta\vartheta}} \left[ \partial_x
  (N \sqrt{q^{xx}} q_{\vartheta\vartheta} \partial_x \phi )+ N \sqrt{q_{xx}}
  \left( \bar{\Delta} \phi - \frac{q_{\vartheta\vartheta}}{2} \frac{\partial
    V}{\partial \phi} \right) \right].
\label{eq:2nd}
\end{equation}
This expression is identical to the wave equation for a massive scalar field 
in curved spacetime as obtain in Appendix~\ref{sec:appD}.

We may further simplify~\eqref{eq:2nd} by taking $q_{\vartheta\vartheta} = x^2$
to obtain 

\begin{equation}
\ddot{\phi} = N \sqrt{q^{xx}} \left[ N \sqrt{q^{xx}} \partial_x^2 \phi +
  \frac{1}{x^2} \partial_x 
  (N \sqrt{q^{xx}} x^2 ) \partial_x \phi + N \sqrt{q_{xx}}
  \left( \Delta \phi - \frac{1}{2} \frac{\partial
    V}{\partial \phi} \right) \right],
\end{equation}
where $\bar{\Delta}^\varphi = \Delta^\varphi /x^2$.

In the classical limit,

\begin{equation}
  N = \sqrt{1 - \frac{2M}{x}} \quad\text{and}\quad
  q^{xx} = 1 - \frac{2M}{x},
\end{equation}
which gives

\begin{equation}
\ddot{\phi} = \left( 1 - \frac{2M}{x} \right) \left[ \left( 1 - \frac{2M}{x}
\right) \partial_x^2\phi + \frac{2}{x} \left(1 - \frac{M}{x} \right)
\partial_x\phi + \Delta\phi -\frac{\partial V}{\partial\phi} \right].  
\end{equation}
This is the well known Klein-Gordon equation in curved Schwarzschild
spacetime~\cite{Bojowald_2010}.

\subsection{Dust Hamiltonian}

In this appendix, we derive the Hamiltonian for
dust~\cite{Brown:1994py,Husain:2011tk}.  
The Lagrangian density for dust is

\begin{equation}
\mathcal{L}_\mathrm{m} = -\frac{1}{2} \sqrt{-g} \left[ g^{\mu\nu} M \left(
    \partial_\mu T \partial_\nu T + 1\right) \right],
\end{equation}
where $T$ is the dust field.
When the dust field is defined in terms of the four-velocity $U_a = \partial_a
T$, $M$ is the rest mass in the stress-energy tensor.

Using the full ADM form of the metric, the Lagrangian density becomes 

\begin{equation}
  \mathcal{L}_\mathrm{m} = -\frac{1}{2} M N\sqrt{q} \left[ -\frac{1}{N^2}
      \left(\dot{T}\right)^2 + 2 \frac{N^a}{N^2} \dot{T} \partial_a T +
      \left( q^{ab} - \frac{N^aN^b}{N^2} \right) \partial_a T \partial_b
      T + 1\right]. 
\end{equation}

The conjugate momentum of the dust field is

\begin{equation}
P_T = \frac{\delta\mathcal{L}_\mathrm{m}}{\delta \dot{T}} =
\frac{\sqrt{q}M}{N} \left( \dot{T} - N^a \partial_a T\right).
\end{equation}

After some algebra similar to above, the constraints are

\begin{align}
  \mathcal{H}_a^\mathrm{m} &= P_T \partial_x T,\\
  \mathcal{H}^\mathrm{m}   &= \frac{1}{2} \left[ \frac{P_T^2}{M\sqrt{q}} +
  M\sqrt{q} \left( q^{ab} \partial_a T \partial_b T + 1
  \right) \right].
\end{align}

The equation of motion for $M$ is

\begin{equation}
\frac{\delta\mathcal{L}_\mathrm{m}}{\delta M} = -N
\frac{\partial\mathcal{H}}{\partial M} = 0  \quad\Rightarrow\quad M =
\frac{P_T}{\sqrt{q}} \left[ q^{ab} \partial_a T \partial_b T + 1
  \right]^{-1/2}.  
\end{equation}

Substituting $M$ back into $\mathcal{H}^\mathrm{m}$ gives

\begin{equation}
\mathcal{H}^\mathrm{m} = P_T \sqrt{q^{ab} \partial_a T \partial_b T + 1}.
\end{equation}

\section{Equation of motion for scalars}\label{sec:appD}

In this appendix, we calculate the equation of motion for a scalar field on a
curved spacetime background.
The wave equation for a massive scalar field is

\begin{equation}
(\Box - m^2)\phi = 0,
\end{equation}
where $\phi = \phi(r,\vartheta,\varphi,t)$ and $m$ is the mass of the field.

In curved spacetime,

\begin{equation}
\left(\nabla_\mu \nabla^\mu -m^2 \right)\phi = 0.
\end{equation}
Since $\nabla^\mu\phi = \partial^\mu\phi$ for a scalar,

\begin{equation}
\left(\nabla_\mu \partial^\mu - m^2\right)\phi = 0.
\end{equation}
Using the identity

\begin{equation}
\nabla_\mu \partial^\mu \phi = \frac{1}{\sqrt{-g}}
\partial_\mu \left( \sqrt{-g} \partial^\mu\phi \right)
\end{equation}
we obtain

\begin{equation}
\frac{1}{\sqrt{-g}} \partial_\mu (\sqrt{-g}\partial^\mu \phi) - m^2\phi = 
\frac{1}{\sqrt{-g}} \partial_\mu (g^{\mu\nu} \sqrt{-g} \partial_\nu \phi) -
m^2\phi = 0.
\end{equation}

For a spherically symmetric static diagonal line element, the metric functions
can be written as 

\begin{equation}
g_{tt}(r), \quad g_{rr}(r), \quad g_{\vartheta\vartheta}(r),
\quad\text{and}\quad 
g_{\varphi\varphi}(r) = g_{\vartheta\vartheta}(r)\sin^2\vartheta. 
\end{equation}
The inverse metric components are

\begin{equation}
g^{tt}(r) = \frac{1}{g_{tt}(r)}, \quad g^{rr}(r) = g_{rr}(r), \quad
g^{\vartheta\vartheta}(r) = \frac{1}{g_{\vartheta\vartheta}(r)},
\quad\text{and}\quad g^{\varphi\varphi}(r) =
\frac{1}{g_{\vartheta\vartheta}(r)\sin^2\vartheta}. 
\end{equation}
From now on the dependence on $r$ will be implicit.

The determinant is

\begin{equation}
\sqrt{-g} = \sqrt{-g_{tt} g_{rr}} g_{\vartheta\vartheta} \sin\vartheta\, .
\end{equation}

Evaluating each component of the wave equation gives

\begin{eqnarray}
\partial_t (g^{tt}\sqrt{-g}\partial_t\phi)
& = & \frac{1}{g_{tt}} \sqrt{-g} \frac{\partial^2\phi}{\partial t^2}
\to \frac{1}{g_{tt}} \frac{\partial^2\phi}{\partial t^2},\\ 
\partial_r (g^{rr}\sqrt{-g}\partial_r\phi)
& = & \frac{\partial}{\partial r} \left( \frac{1}{g_{rr}} \sqrt{-g_{tt}
g_{rr}} g_{\vartheta\vartheta} \sin\vartheta \frac{\partial\phi}{\partial r}
\right) = \sin\vartheta \frac{\partial}{\partial r} \left(
\sqrt{\frac{-g_{tt}}{g_{rr}}} g_{\vartheta\vartheta} \frac{\partial
\phi}{\partial r} \right)\nonumber\\ 
& \to & \frac{1}{\sqrt{-g_{tt} g_{rr}}g_{\vartheta\vartheta}}
\frac{\partial}{\partial r} \left( \sqrt{ \frac{-g_{tt}} {g_{rr}} }
g_{\vartheta\vartheta}  \frac{\partial \phi}{\partial r} \right),\\
\partial_\vartheta \left( g^{\vartheta\vartheta} \sqrt{-g}\partial_\vartheta \phi
\right)
& = & \frac{1}{g_{\vartheta\vartheta}} \frac{\partial}{\partial\vartheta} \left(
\sqrt{-g_{tt} g_{rr}} g_{\vartheta\vartheta} \sin\vartheta
\frac{\partial\phi}{\partial\vartheta} \right)
= \sqrt{-g_{tt} g_{rr}} \frac{\partial}{\partial \vartheta} \left(
\sin\vartheta \frac{\partial\phi}{\partial\vartheta} \right)\nonumber\\
& \to & \frac{1}{g_{\vartheta\vartheta}\sin\vartheta}
\frac{\partial}{\partial\vartheta} \left( \sin\vartheta
\frac{\partial \phi}{\partial \vartheta} \right),\\
\partial_\varphi (g^{\varphi\varphi}\sqrt{-g}\partial_\varphi\phi)
& = & \frac{1}{g_{\vartheta\vartheta}\sin^2\vartheta}\sqrt{-g}
\frac{\partial^2\phi}{\partial \varphi^2}
\to \frac{1}{g_{\vartheta\vartheta}\sin^2\vartheta}
\frac{\partial^2 \phi}{\partial \varphi^2},\\
& & -\sqrt{-g} m^2 \phi \to -m^2\phi.
\end{eqnarray}
The arrow indicates that the expression has been  divided by $\sqrt{-g}$.

The wave equation becomes

\begin{equation}
\frac{1}{g_{tt}} \frac{\partial^2 \phi}{\partial t^2}
+ \frac{1}{ \sqrt{ -g_{tt} g_{rr} } g_{\vartheta\vartheta} }
\frac{\partial}{\partial r} \left( \sqrt{\frac{-g_{tt}}{g_{rr}}} 
g_{\vartheta\vartheta} \frac{\partial \phi}{\partial r} \right)
+ \frac{1}{g_{\vartheta\vartheta}} \Delta^\varphi \phi
- m^2 \phi = 0,
\end{equation}
where 
\begin{equation}
\Delta^\varphi = 
  \frac{1}{\sin\vartheta} 
  \frac{\partial}{\partial\vartheta} \left( \sin\vartheta \frac{\partial
    \phi}{\partial\vartheta}  \right)
+ \frac{1}{\sin^2\vartheta} \frac{\partial^2 \phi}{\partial\vartheta^2}
\end{equation}
is the Laplacian on a two-sphere.
This form of the scalar wave equation corresponds to motion in a general
spherically symmetric static spacetime. 

\section*{Acknowledgments}
I thank Saeed Rastgoo for helpful discussions.
We acknowledge the support of the Natural Sciences and Engineering
Research Council of Canada (NSERC). 
Nous remercions le Conseil de recherches en sciences naturelles et en
g{\'e}nie du Canada (CRSNG) de son soutien. 
\bibliographystyle{JHEP}
\bibliography{gingrich}
\end{document}